\documentstyle[epsf,psfig,amstex,amssymb]{l-aa}

\begin{document}
\thesaurus{ XX }        

\title{Non-linear Dynamics and Mass Function  of Cosmic Structures:  I
  Analytical   Results} \author{Edouard Audit$^1$, Romain Teyssier$^2$
  and   Jean-Michel     Alimi$^1$}     \institute{$^1$     Laboratoire
  d'Astrophysique  Extragalactique et de  Cosmologie, Observatoire  de
  Paris-Meudon, Meudon 92195 Cedex,
  France\\
  $^2$  CEA,  DSM/DAPNIA/Service d'Astrophysique  CE-SACLAY,   F-91191
  Gif--sur--Yvette,   France}  
\date{} 
\maketitle 
\markboth{Non-linear  Dynamics and Mass Function} {I Analytical Results}
% my definitions
\def\etal{{\it et al. \/}}
\def\eg{{e.g.,\ }}
\def\etc{{etc.\ }}
\def\ie{{i.e.,\ }}
\unitlength=1cm

\begin{abstract}
  We investigate some modifications  to the Press \& Schechter  (1974)
  (PS) prescription resulting   from  shear and tidal  effects.  These
  modifications rely   on more realistic  treatments   of the collapse
  process than the standard approach based on the spherical model.
  
  First, we  show   that the  mass  function   resulting from  a   new
  approximate Lagrangian dynamic  (\cite{AA96}), contains more objects
  at high mass, than the classical PS mass function and is well fitted
  by a PS-like  function with a threshold  density of $\delta_c \simeq
  1.4$. However, such  a Lagrangian description can  underestimate the
  epoch of structure formation since it  defines it as the collapse of
  the first  principal  axis.   We therefore suggest   some analytical
  prescriptions, for computing the collapse  time along the second and
  third   principal   axes,  and  we  deduce    the corresponding mass
  functions. The collapse along the third axis is delayed by the shear
  and the number of objects of high mass then decreases.
  
  Finally, we show that the shear  also strongly affects the formation
  of low-mass  halos. This dynamical effect  implies a modification of
  the low-mass slope of the mass function  and allows the reproduction
  of  the observed   luminosity function  of  field  galaxies.  In   a
  companion paper, we  present results of numerical simulations  which
  complete this work.
\end{abstract}
\section{INTRODUCTION.}

One of the simplest   statistical indicators which  characterizes  the
mass  distribution  in the universe  is   the mass function  of cosmic
structures.  Press and Schechter (1974, hereafter PS), have proposed a
formalism to  evaluate  this  function  at  any   time.  For  a  given
hierarchical scenario of structure formation, as for instance CDM, the
mass function is obtained using only the power spectrum of the initial
density fluctuations.  This formalism has been extensively tested with
the help  of numerical simulations.  The  first comparison between the
mass function of  dark-matter halos obtained  in numerical simulations
and that predicted in the  framework of the  PS approach have shown  a
satisfactory  agreement  (\cite{EFWD88}).   This  agreement was  later
improved by imposing   some  ad--hoc modifications to the   initial PS
formalism (\cite{ER88},  \cite{CC89},  \cite{LC94}).  For  example, it
was shown in these studies  that the concordance between the numerical
and the  analytical predictions was  improved by lowering  the initial
density threshold  at which a spherical  homogeneous region collapses. 
However, the  theoretical reasons for which  a quantity resulting from
very non--linear dynamics, such as  the mass function, can be computed
from this formalism have never been clarified.  This question takes on
a greater  importance because the   PS formalism which is  a  powerful
tool, is widely used in phenomenological analytical cosmology.  It is,
for example, very often used  to put  constraints on the  cosmological
parameters  using   either  the   luminosity   function  of   galaxies
(\cite{WK91}, \cite{BVGM92}), the  temperature distribution  of  X-ray
clusters  (\cite{OB92}, \cite{Via96}) or the Sunayev-Zel'dovich effect
(\cite{CK88}, \cite{BBBO96}, \cite{Eke96} and references therein).

The  mass function of cosmic  structures  obtained in the framework of
the PS formalism has  a {\bf statistical} and  {\bf dynamical} origin. 
Starting with the statistical properties of the initial density field,
PS identify the fraction of mass embedded in collapsed objects of mass
greater than  a given mass M, with  the fraction of volume filled with
regions of space satisfying a given  dynamical criterion.  Usually the
criterion is  simply issued from the  spherical collapse which selects
all the points whose initial density  is above a certain threshold. It
is highly probable that such a simple model  is unable to describe the
complex features   occurring during   a  fully non--linear   dynamical
process.   Therefore, one can imagine  that only a fraction of regions
with a given  density contrast   exceeding the threshold    eventually
collapses.  This ``fuzzy'' density  criterion    can be viewed  as   a
statistical tool (\cite{BVGM92}), but here, in  order to have a better
understanding of structure formation, we intend  to derive it from the
gravitational dynamics.

Therefore, in this   paper,   we study  the  influence of   non-linear
dynamics on the  mass function, highlighting  the role played by shear
and  tidal effects  on the  actual  shape of  the  mass  function.  We
present analytical results which  exhibit generic features of PS--like
mass functions and  show the relationship  between the  dynamics and the
mass  function through what  we call a  selection function.  Our study
concerns both  non-spherical  Lagrangian  dynamical  models, and   some
dynamical  ``ansatz'' that we   introduce  to describe the   structure
formation  beyond     the  usual    validity  range     of  Lagrangian
approximations.    In  a  companion paper   we compare  our analytical
predictions with numerical results.

%%%%%%%%%%%%%%%%%%%%%%%%%%%%%%%%%%%%%%%%%%%%%%%%%
\section{Mass Functions from Lagrangian Dynamics}
%%%%%%%%%%%%%%%%%%%%%%%%%%%%%%%%%%%%%%%%%%%%%%%%%

\subsection{PS formalism from a Lagrangian point of view.}

The analytical method  which  has been most successful  up  to  now to
compute the mass function is the one introduced  by PS.  In this paper
we use such a  formalism but we intend  to give our own interpretation
of  this approach   from  a  Lagrangian   point of  view    (see  also
\cite{M95}). This gives us a  general framework for computing the mass
function.

The  idea  of  PS is  to  compute the  number density  of cosmological
objects with mass $M$ at a given time,  by identifying the fraction of
mass contained in these  objects with the  mass embedded in the  space
filled by points  with  a density  which reaches infinity  at the same
time.  When  considering an object of   mass $M$, the  initial density
field is smoothed on a  scale $R$ with  $M  \propto R^{3}$. To give  a
concrete expression for this  idea one needs a  dynamical prescription
that allows  to  follow the  evolution  of each point of   the initial
density field and to know its statistical properties.

Throughout this paper  we work in a Friedman-Robertson-Walker universe
with a critical density parameter,  $\Omega = 1$, and initial Gaussian
random  fluctuations.   At   some early  epoch,    corresponding to an
expansion    factor $a_{i}$  and     a  redshift $z_i$, the    density
fluctuations are entirely described by their power spectrum.  We first
multiply this spectrum with  a low-pass filter of characteristic scale
$R$ that cuts the power above a frequency $\simeq 2\pi/R$.  The choice
of filter  is,    a  priori,  arbitrary~(\cite{LC94}).   We   restrict
ourselves to  the Top--Hat window function  in real space, because its
application has  an  intuitive physical interpretation.   It allows to
consider homogeneous  and spherical regions  of space.  We then assume
that the resulting density field is correctly described by independent
fluid elements.  Each fluid element is now initially described by some
dynamical   parameters,   which    are  obtained  by    smoothing  the
corresponding   fields  to  the   scale    $R$.  Given  these  initial
conditions, we evolve each fluid element separately until it collapses
(i.e. until the density of  the fluid element  becomes infinite).  The
evolution has  to be followed in  a Lagrangian way  in order to relate
the   {\bf   collapse epoch} of   the  fluid  element  to  its initial
conditions.  In the standard  PS approach, these fluid elements evolve
according to the  homogeneous spherical   model.   In this case,   the
collapse epoch is entirely  determined by the initial density contrast
of the fluid  element.  However, it  has been shown by several authors
(e.g.    \cite{vdW94}, \cite{AA96})  that,   during the  gravitational
collapse,   the  shear  and  the  tidal forces    are  very important. 
Consequently, more complex, non-spherical, gravitational dynamics have
to be considered  in order to  take into consideration these  physical
effects which are totally discarded in the spherical model.

For Lagrangian non-spherical dynamics, the   collapse time $a_c$ of  a
fluid  element   can depend on   several   initial smoothed parameters
$x^{j}$, $j=1,n$. The fraction of mass embedded  at present ($a=1$) in
collapsed structures with a mass greater or equal to $M$ is then given
by

\begin{equation}
F(\ge M) = \int 
             P_R\left(x^{j}\right) s\left(x^{j}\right)
             dx^{1} \cdot \cdot  \cdot dx^{n}
\label{fraction}
\end{equation}

\noindent
where $s=1$ if $a_c\left(x^{j}\right)  \le 1$ and  $0$ otherwise.  The
selection  function  $s$ makes  a  direct link  between the Lagrangian
dynamics of the fluid elements and the  mass function. The integral of
$s$, with respect to all  $x_j$ parameters except the initial  density
contrast, can be interpreted as a  statistical quantity. This quantity
gives the probability that  a given region  of space becomes part of a
bound   structure     according to   its    initial  density  contrast
(\cite{BVGM92}).  The   right-hand side  of  equation (\ref{fraction})
depends  on $M$   through the  smoothing  scale   $R$.   $P_R$ is  the
probability  density function of the   initial smoothed parameters, it
represents the fraction  of volume   filled by  the regions  of  space
having initial parameters $(x^{j}$, $j=1,n)$.

In   general, the    above definition  runs    into   the so    called
``renormalization problem'', i.e.   $F(\ge 0)=F_0\ne 1$.  This problem
can be alleviated if one carefully considers the fate of regions which
have not been  selected  on a scale $R$,  but  can be selected with  a
non--zero probability on another scale $R'>R$, and therefore belong to
objects  with   a mass  greater    than  $M$.  Excursion    set theory
(\cite{PH90}, \cite{B91a},  1991b) solves this renormalization problem
for the   particular case  of  the sharp-k  filter, but  the  physical
interpretation of this filter remains unclear.  In this work we do not
investigate this renormalization problem.  We  always assume that this
renormalization  is  justified in  any  case.  We therefore divide the
mass fraction  obtained in equation   (\ref{fraction}) by the constant
$F_0$.  The mass function of collapsed objects, which gives the number
density of object with mass $M$, is then defined as

\begin{equation}
\Phi (M) = - \frac{\rho _{0}}{M} \frac{d}{dM} \left(\frac{F(>M)}{F_{0}}
\label{mfdef}
\right)
\end{equation}

In order to determine the mass function  we therefore have to evaluate
the statistical properties of the initial  parameters and the collapse
epoch associated with these  parameters.   In particular, we  evaluate
the collapse   epoch using  three  different Lagrangian  dynamics: the
spherical  model; the Zel'dovich  approximation  and a  more realistic
local  Lagrangian    model  that we  introduced    in   a former paper
(\cite{AA96}).   These  three dynamical  models  state that  the fluid
element   belongs  to a collapsed    object  when its density  becomes
infinite.   For   the   two  latter   cases,  this  first  singularity
corresponds  to the  collapse of  the  fluid element  along  its first
principal  axis and then favors the  formation of sheet--like objects. 
One can then ask whether such objects agree  with the idea of having a
dense,  virialized  ``blob''   in the   density  field. Unfortunately,
Lagrangian  theory is unable  to  bypass this  first singularity.   We
therefore investigate two other  models   in section 3 which   roughly
describe the collapse of the fluid element  along its second and third
axis.  This  comparative analysis of the  different  mass functions we
obtain for each case  allows one to study  the effect of the dynamics,
and in  particular, the influence  of  the shear and  the  tide on the
formation of structures in a PS formalism.

%%%%%%%%%%%%%%%%%%%%%%%%%%%%%%%%%%%%%%%%%%%%%%
\subsection{Computation of the Collapse Epoch}
%%%%%%%%%%%%%%%%%%%%%%%%%%%%%%%%%%%%%%%%%%%%%%

In the simplest  cases, only   one   initial parameter is  needed   to
determine the collapse epoch of a fluid element.  This is the case for
the  spherical model, where   the  initial density contrast   entirely
determines the evolution of a  fluid  element, and for the  Zel'dovich
approximation where the  smallest eigenvalue of the deformation tensor
gives the collapse epoch.  However, these  dynamics are rather simple. 
It does not seem very realistic to account for the variety of physical
processes   occurring  during  the  collapse  with   a single  initial
parameter.  Therefore,  we also use a  local Lagrangian  dynamic which
takes  into account, to some extend,  the effect of  the shear and the
tide (\cite{AA96}), and then determines the collapse epoch in terms of
the three eigenvalues of the deformation tensor.

The collapse epoch (written here as the  collapse redshift $z_c$) of a
fluid element evolving according to  the  spherical model is given  by
the simple formula

\begin{equation}
1 + z_c = \frac{\delta} {\delta _c}
\end{equation}

\noindent
where $\delta  _{c} = \frac{3}{5}(\frac{3\pi}{2})^{2/3} \simeq 1.686$. 
We use for  all fields its initial value  linearly extrapolated to $z =
0$  (for example:   $\delta  =  (1+z_i)  \delta(z_i)$).    Within  the
spherical model  ,  the collapse epoch depends   only on  the  initial
density contrast and only fluid elements which are initially over-dense,
collapse.

In the framework   of  the Zel'dovich approximation  (\cite{Z70})  the
density contrast of a fluid element evolves according to

\begin{equation}
1+\delta(t) = \frac{1}{(1+a(t)\lambda_{1})(1+a(t)\lambda_{2})
                       (1+a(t)\lambda_{3})}
\label{zeldel}
\end{equation}

\noindent
where  $\lambda_{1}$,  $\lambda_{2}$ and  $\lambda_{3}$  are the three
initial eigenvalues of the  deformation tensor arranged  in decreasing
order.  The collapse epoch of the first principal axis is reached when
the denominator in equation (\ref{zeldel}) vanishes

\begin{equation}
1 + z_c =-\lambda_{3} = \frac{\delta}{3}-\sigma_{3}
\label{zelac}
\end{equation}

\noindent 
Once again $\delta$ is   the  initial  density contrast  of the  fluid
element and   $\sigma_{i}$ are  its initial shear  eigenvalues.  In
the case of  zero shear, the collapse epoch  is  given by:  $1 +  z_c =
\delta / 3$.  This means that the exact solution of the spherical case
is  not recovered, as can be  expected for this  first order dynamical
prescription.  For a fixed density contrast, the shear accelerates the
collapse  ($\sigma_3 \leq  0$)    and equation (\ref{zelac})   can  be
considered as a first simple approximation for non-spherical dynamics.

The  Zel'dovich  approximation illustrates in  a   simple way that the
collapse  epoch cannot  be  considered as a  function  of  the density
contrast  alone.   However,  computing  the  collapse  epoch   in terms  of
$\lambda_{3}$ alone is not satisfactory either.  Therefore, we present
in this paragraph  a  derivation of the  collapse  time  which  uses a
Lagrangian  dynamic based on the deformation  tensor.  Audit \& Alimi
(1996),  hereafter AA, have  proposed  a local Lagrangian dynamic to
determine the evolution of the eigenvalues of  the deformation tensor. 
Knowing these   eigenvalues and their  time  derivatives  is enough to
compute    all   the  other  dynamical    quantities.   The  dynamical
prescription of AA  is well suited to  the computation of the collapse
epoch.  It  accounts  well  for  the effect  of the  shear on the
collapsing fluid element.  It is exact  for spherical, cylindrical and
planar    geometry and very accurate   over  a  wide  range of initial
conditions.   Similar to the  Zel'dovich  approximation, it  predicts that
under-dense regions can collapse due to the effect of the shear.

In terms of the eigenvalues of the deformation tensor, the density
contrast is defined as

$$
1+\delta(\tau)=\frac{1}{ \left(1+\lambda_{1}(\tau)\right)
                         \left(1+\lambda_{2}(\tau)\right)
                         \left(1+\lambda_{3}(\tau)\right)}
$$

The evolution equation for each of the $\lambda_{i}(\tau)$ is given in
the framework of the AA approximation by

\begin{equation}
  \frac{d^{2}\lambda_{i}}{d\tau^{2}} = \left( \frac{1+\lambda_{j}/2
      +\lambda_{k}/2+\lambda_{j}\lambda_{k}/2}
    {1+\lambda_{j}+\lambda_{k}+\lambda_{j}\lambda_{k}} \right)
  \frac{6\lambda_{i}}{\tau ^{2}}
  \label{aadyn}
\end{equation}

\noindent 
where $\tau $ is  the conformal time  defined by: $ d\tau  = dt / a^2$
and $(i,j,k)$ is a   circular permutation of $(1,2,3)$.    These three
equations are highly non linear and have been integrated numerically.

The initial  eigenvalues  of the   shear  and deformation tensors  are
related by:   $\lambda_{i}=\sigma_{i} - \delta/3$.   The two triplets
$(\lambda_{1}, \lambda_{2},  \lambda_{3})$ and  $(\delta,  \sigma_{2},
\sigma_{3})$  are  therefore  totally  equivalent    ($\sigma_{1}$  is
obtained    using   the  relation:   $\sigma_{1}     + \sigma_{2}  +
\sigma_{3}=0$).   In  the following, we   prefer  to  use the  triplet
$(\delta, \sigma_{2}, \sigma_{3})$, which allows an immediate physical
interpretation. The  eigenvalues of the  shear  and deformation tensor
are always arranged in decreasing  order: $\sigma_{1} \geq \sigma_{2}
\geq \sigma_{3}$ and $\lambda_{1} \geq \lambda_{2} \geq \lambda_{3} $.

Moreover,  in order to integrate  equation (\ref{aadyn})  we only keep
the  linear growing mode in the  initial conditions.  This restriction
and the invariance of equation (\ref{aadyn}) under the transformation:
$\tau \rightarrow  \alpha \tau$ ensures  the following scaling law for
the collapse epoch

\begin{equation}
a_c(\delta,\sigma_{2},\sigma_{3}) =
\alpha a_c(\alpha\delta,\alpha\sigma_{2},\alpha\sigma_{3})
\label{scalinglaw}
\end{equation}

\begin{figure}[httb]
\begin{picture}(9,9)
\put(-0.5,0){
\psfig{file=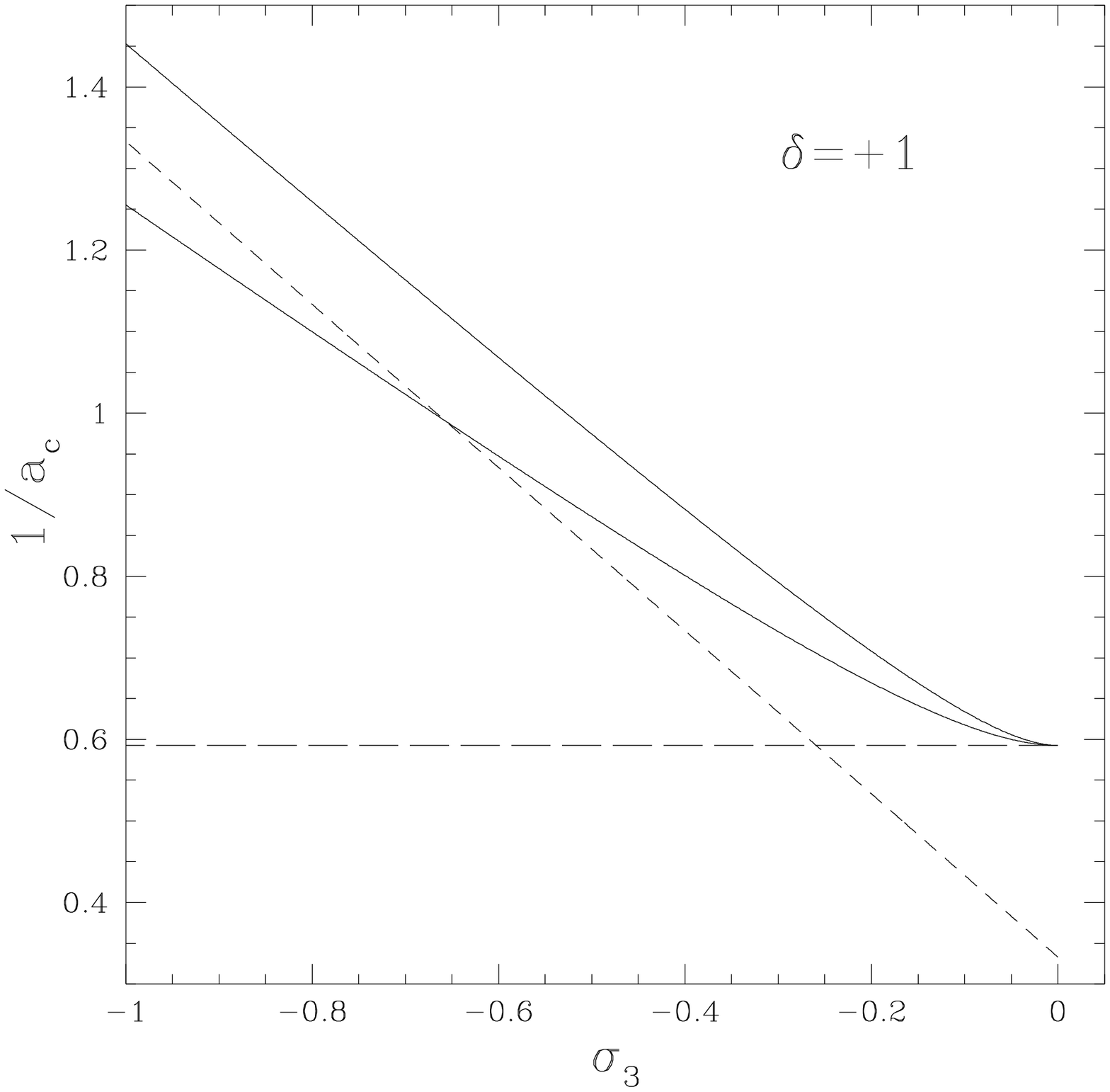,height=4.5cm,width=4.5cm}}
\put(4.,0){
\psfig{file=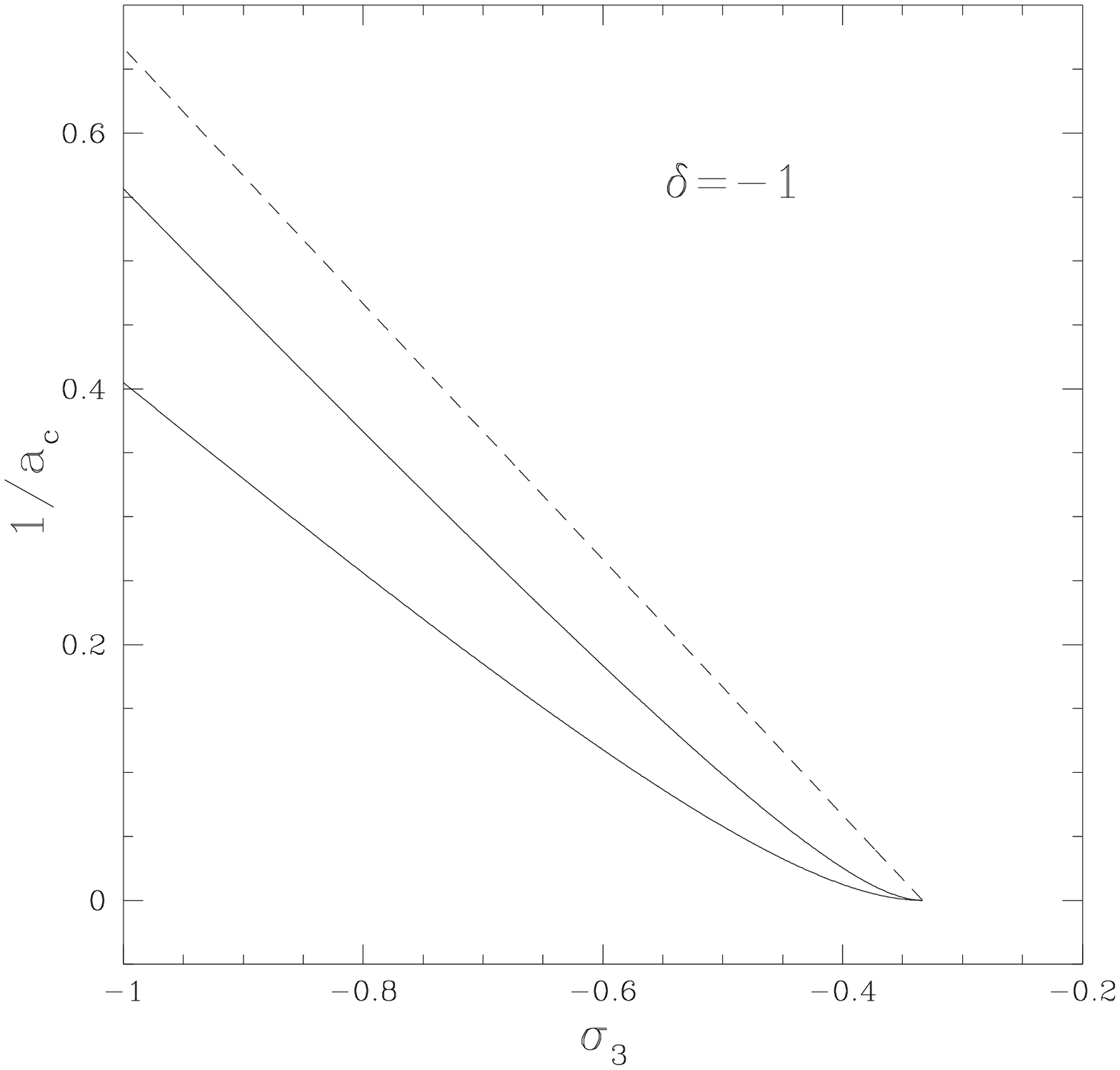,height=4.5cm,width=4.5cm}}
\put(-0.5,4.5){
\psfig{file=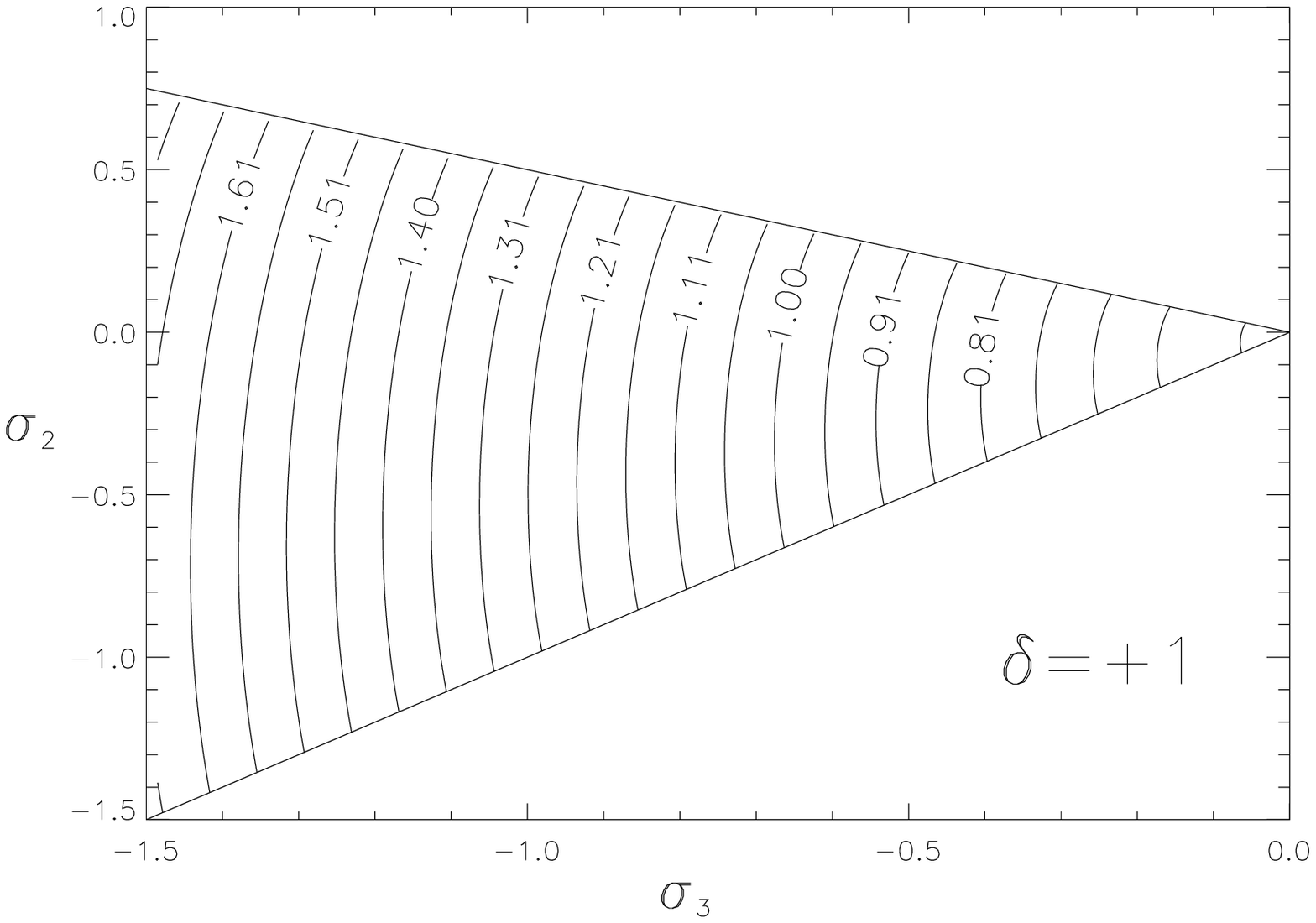,height=4.5cm,width=4.5cm}}
\put(4.,4.5){
\psfig{file=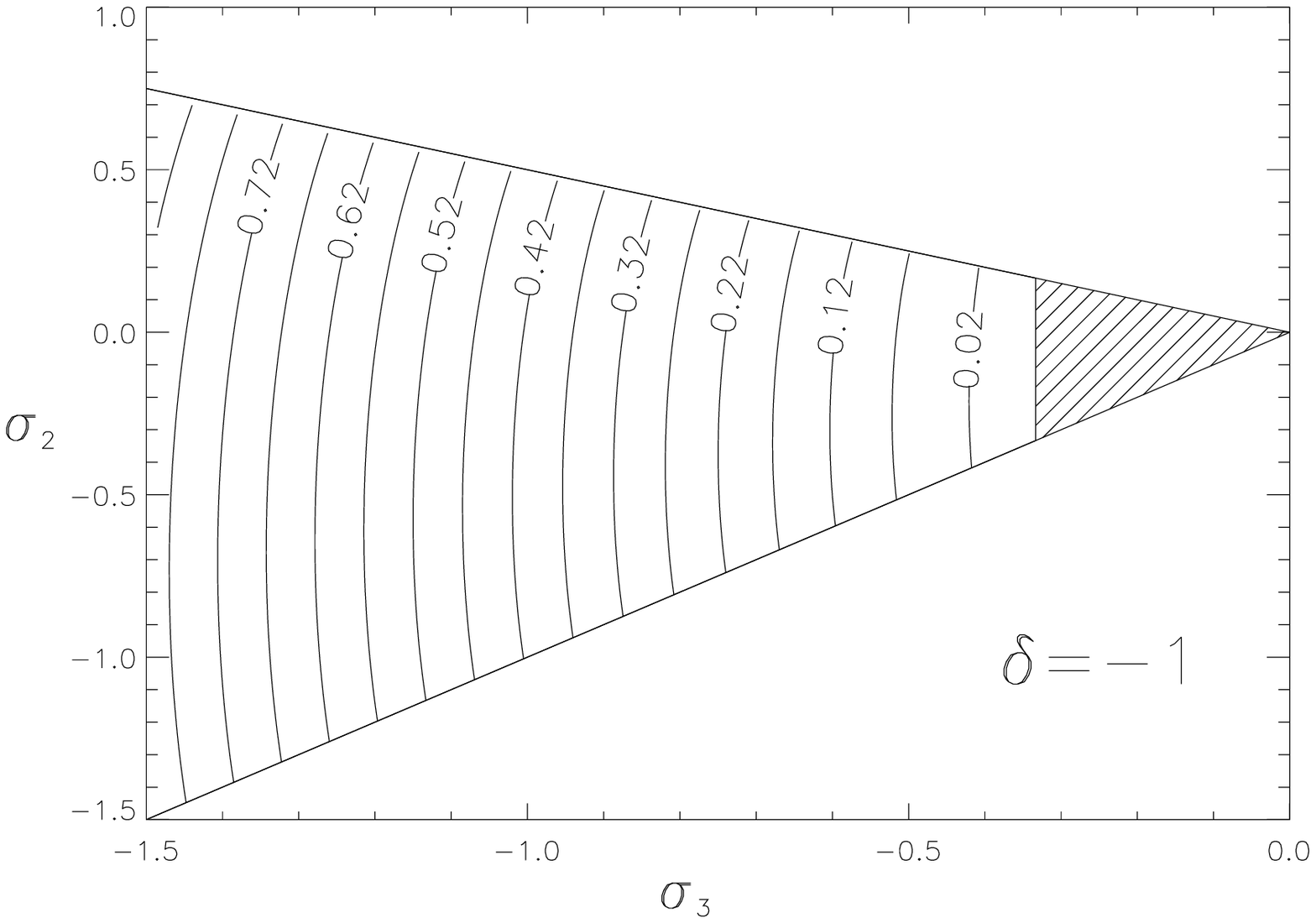,height=4.5cm,width=4.5cm}}
\end{picture}
\caption{Upper panels: Iso-contours of the redshift of collapse in the
  ($\sigma_3,~\sigma_2$) plane  for $\delta=+1$ and  $\delta=-1$.  The
  shaded area  corresponds to  fluid  elements  which never collapse.  
  Lower panels: The collapse redshift (or the  inverse of the collapse
  epoch)  as a function of the  smallest eigenvalue of  the shear. The
  solid  curves correspond to  the two boundary  lines of the triangles
  plotted in the upper graphs.   The small-dashed lines correspond  to
  the Zel'dovich approximation. For $\delta=+1$ the spherical model is
  shown as a long-dashed line.}
\label{invtc}
\end{figure}

Such a property of the collapse time is also present for the spherical
model and the  Zel'dovich  approximation. It allows  to reduce  by one
dimension the parameter space that we have to explore to determine the
collapse epoch for any initial  condition.  It is therefore sufficient
to  know the function   ${\cal  T}^{+}(  \sigma_2, \sigma_3)$ (resp.   
${\cal T}^{-} (\sigma_2, \sigma_3)$)  representing the inverse  of the
collapse epoch ($a_c^{-1}=1+z_c$)  in the $(\sigma_2, \sigma_3)$ plane
for $\delta =  +1$ (resp.  $\delta =  -1$).   Then, we can deduce  the
collapse epoch for any set of initial conditions using the scaling law
of equation (\ref{scalinglaw}).   In figure (\ref{invtc}), we plot the
iso-contours of the functions ${\cal T}^{+}$ and ${\cal T}^{-}$.  From
the previously given relations between  the $\sigma_{i}$ it is easy to
deduce that $\sigma_{3}  \leq  \sigma_{2} \leq -  \sigma_{3}/2$.  This
explains why only  a portion of the  $( \sigma_2, \sigma_3 )$ plane is
covered.  For each of these plots, a thousand  points were computed by
numerically solving equation  (\ref{aadyn}).  We present an analytical
fit (accurate to within  $1\%$) of these two  surfaces in Appendix~A.  
For $ \delta  >  0$, the fluid element   always collapses in a  finite
time.  For  $\delta < 0$,  a finite area of  shear values  is singular
(i.e.  the  collapse  time is infinite).  This   area corresponds to a
fluid  element whose initial   deformation  tensor has three  positive
eigenvalues (all fluid elements  with at least one negative eigenvalue
collapse).

The two lower panels  of figure [\ref{invtc}] show  the inverse of the
collapse  time  as  a   function  of $\sigma_3$   for $\delta=+1$  and
$\delta=-1$.  The long--dashed line corresponds to the spherical model
for which  the collapse    time is  independent  of   $\sigma_3$.  The
small--dashed line corresponds to  the Zel'dovich approximation.   The
full lines represent  the inverse  collapse time of  the AA  dynamics. 
The upper line corresponds  to the case where $\sigma_2=\sigma_3$ (the
fastest collapse at given $\sigma_3$ and $\delta$)  and the lower line
to  $\sigma_2=-\sigma_3/2$ (the  slowest  collapse at given $\sigma_3$
and $\delta$).  For $\delta < 0$ the collapse time becomes infinite as
$\sigma_3$ reaches $\delta/3$.  The spread  between the two full lines
gives an idea of the  influence of $\sigma_2$ on  the dynamics and  of
the  importance of the coupling  between  the different directions  of
collapse of the fluid element.  Note  however that the error one could
make by neglecting the influence of $\sigma_2$ to compute the collapse
epoch of the first principal axis is of the order of 20\%.

%%%%%%%%%%%%%%%%%%%%%%%%%%%%%%%%%%%%%%%%%%%%%%%%%
\subsection{Statistics of the Initial Parameters}
%%%%%%%%%%%%%%%%%%%%%%%%%%%%%%%%%%%%%%%%%%%%%%%%%

In the  case of initial  Gaussian    density fluctuations and  assuming
an initial  linear growth  mode  only, Doroskevitch  (1970)  derived the
probability function for the eigenvalues of the deformation tensor, namely

\begin{eqnarray}
& &P(\lambda_{1},\lambda _{2},\lambda _{3})=
\frac{675\sqrt 5}{8\pi \Delta ^{6}}
e^{-\frac{3s_{1}^{2}}{\Delta ^{2}}+\frac{15s_{2}}{2\Delta^{2}}}\nonumber \\ 
& &\hspace{2.5cm} \times (\lambda_{1}-\lambda_{2})(\lambda_{1}-\lambda_{3})
(\lambda_{2}-\lambda_{3})
\label{doro}
\end{eqnarray}

\noindent 
where $\Delta$ is the r.m.s density contrast smoothed at scale $R$

\begin{equation}
\Delta ^{2} (R) = \int 4\pi k^{2}P(k)W^{2}(kR)dk
\end{equation} 

\noindent 
and $ s_1 = \lambda _{1}  + \lambda _{2} + \lambda  _{3} $ and $ s_2 =
\lambda _{1} \lambda _{2} +  \lambda _{2} \lambda  _{3} + \lambda _{1}
\lambda _{3} $.    The $\lambda$'s are correlated  variables,
$\delta$ and the  doublet $(\sigma_2,\sigma_3)$ are  independent.  The
probability distribution  for these new  variables can then be deduced
easily from equation (\ref{doro})

\begin{equation} 
P(\delta, \sigma _{2}, \sigma _{3}) = \frac{1}{\Delta^3}
P_{\nu}\left(\frac{\delta}{\Delta}\right)
P_{\mu_2,\mu_3}\left(\frac{\sigma_{2}}{\Delta},
                         \frac{\sigma_{3}}{\Delta}\right)
\end{equation} 

\noindent 
where the  marginal  probability functions for the  rescaled variables
$\nu = \delta / \Delta$ and $\mu_i = \sigma_i / \Delta$ are given by

\begin{equation} 
P_{\nu} = \frac{1}{\sqrt{2\pi}}e^{-\nu ^{2}/2}
\end{equation}

\noindent 
and 

\begin{eqnarray}
& &P_{\mu _{2},\mu _{3}} 
                           = \frac{675}{4}\sqrt {\frac{5}{2\pi}}
e^{-7.5(\mu_{2}^{2}+\mu_{2}\mu_{3}+\mu_{3}^{2})} \nonumber \\
& &\hspace{2cm}\times (\mu_{2}+2\mu_{3})(\mu_{3}+2\mu_{2})(\mu_{2}-\mu_{3}) 
\nonumber
\end{eqnarray}

\noindent 
We compute  also  the marginal probability   distribution functions of
$\mu _{1} \in [0, +\infty[$ and $\mu_{3} \in ]-\infty, 0]$

\begin{eqnarray}
& &P_{\mu_1} = P_{\mu_3}
= 3\sqrt{\frac{5}{2\pi}}e^{-45\mu_3^{2} / 8}\\ 
& &\hspace{2cm}\times \left(e^{-135\mu_3 ^{2}/8} 
                          -(1-\frac{135}{8}\mu_3^{2}) \right)
\nonumber
\end{eqnarray}

\noindent
and of $\mu_{2} \in  ]-\infty, +\infty[$

\begin{equation}
P_{\mu_2}=3\sqrt{\frac{5}{2\pi}}e^{-45\mu_2^{2} / 2}
\end{equation}

Equation (\ref{fraction}) can be now  re-written in terms of $\nu$,
$\mu_{2}$, $\mu_{3}$ and $\Delta$.

\begin{equation}
\!\!\!\!\!\!
F(\ge M) = \!\! \int \!\! 
           P_{\nu}P_{\mu _{2},\mu _{3}}
           s(\nu\Delta,\mu_2\Delta,\mu_3\Delta)
           d\nu d\mu_2 d\mu_3
\label{funi} 
\end{equation}

The only quantity which then depends on $M$ is $\Delta$. Consequently,
the mass function can be expressed as

\begin{eqnarray}
\Phi(M) & = &-\frac{\rho_0}{M} \frac{d\Delta}{dM}\frac{1}{F_0} \nonumber \\  
&\times  &    \int  P_{\nu}P_{\mu _{2},\mu _{3}}
        \frac{\partial s}{\partial \Delta}
        (\nu\Delta,\mu_2\Delta,\mu_3\Delta)
        d\nu d\mu_2 d\mu_3
\end{eqnarray}

\noindent 
where the integral is a  function of $\Delta$  and depends only on the
chosen dynamics which gives its specific form to  $s$. It is therefore
possible  to define a universal mass  function for each dynamic which
is    independent  of  the initial   power  spectrum  of   the density
fluctuation.  The mass function can  be written, for any spectrum, in
the form

\begin{equation} 
\Phi (M)=-\frac{\rho_0}{M}\frac{d\Delta}{dM} \Phi(\Delta)
\label{scaling}
\end{equation} 

The spectrum dependence  of the mass function  is totally contained in
the $ d\Delta/dM$ factor.   We call $\Phi(\Delta)$, which depends only
on the chosen dynamics  (i.e.  on the  function $1  + z_c$),  the {\bf
  universal mass function}.   This demonstrates that the mass function
resulting from  a PS  formalism admits, in  addition to  the well know
time   self-similarity, a ``scaling'' behavior  in  spectrum (see also
\cite{LC94}).   This property   holds only  in  a  critical  universe,
independently of the dynamics as long as only the growing mode is kept
during  the  linear regime.  These  two  ``scaling properties'' of the
mass function are general characteristics of a PS--like mass function.
Therefore, they can be used as test for the validity  of a PS approach
to the mass function.

%%%%%%%%%%%%%%%%%%%%%%%%%%%
\subsection{Mass Functions}
%%%%%%%%%%%%%%%%%%%%%%%%%%%

Equation (\ref{fraction}) can now be re-written as follows

\begin{equation} 
F(\ge M) = \frac{1}{\sqrt{2\pi}} \int _{-\infty} ^{+\infty}
e^{-\nu^{2}/2} S(\nu\Delta,\Delta )d\nu
\label{fraction2}
\end{equation}

\noindent 
For the spherical  model, $S$ is identical  to  the selection function
$s$ which depends then only on the initial density contrast

\begin{equation} 
s(\nu \Delta) = \left\{ \begin{array}{l}
                     0 \mbox{ if } \nu \Delta < \delta_c \\
                     1 \mbox{ otherwise } \\
                     \end{array}
                     \right.
\end{equation}

\begin{figure}[httb]
\psfig{file=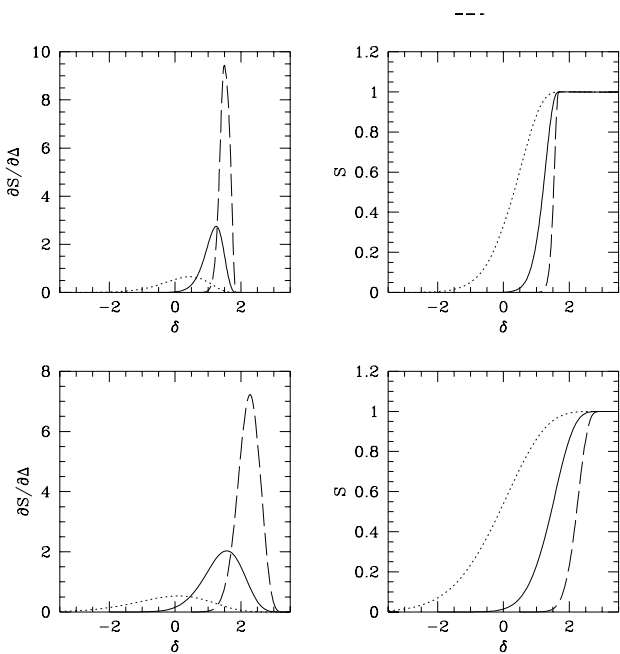,height=9cm}
\caption{Selection functions and their derivatives (see text for
  definitions) obtained at $\Delta$=2 (dotted line), $\Delta$=1 (solid
  line) and  $\Delta$=0.5 (dashed line) for  the Zel'dovich model (two
  lower graphs) and for the AA model (two upper graphs). }
\label{selfunc}
\end{figure}

For the  Zel'dovich and AA approximations,  the selection  function is
not   a  function  of   the   density contrast  alone.    In  equation
(\ref{fraction2}), $S$ is then the selection function, integrated over
all parameters ($\mu_3$ for Zel'dovich  approximation and $\mu_2$  and
$\mu_3$  for AA approximation), except  $\nu$.   For the AA  dynamical
model, $S$ is explicitly given by
\begin{equation}
\begin{array}{ll}
S(\nu\Delta,\Delta)= \int_{-\infty}^{0} \int_{\mu_3}^{-\mu_3/2} 
                   &S(\nu \Delta,\mu_2 \Delta,\mu_3 \Delta)\\
                   &\times P_{\mu_{2},\mu_{3}}d\mu_{2}d\mu_{3}  \\
\end{array}
\end{equation} 
It  represents  the  probability  that a  fluid  element  with a given
initial  density contrast, collapses  in the age of  the Universe.  In
figure (\ref{selfunc}), we plot this quantity  and its derivative with
respect to $\Delta$ for $\Delta$   = 0.5, 1 and   2. A small  $\Delta$
value corresponds to high mass objects, and  a large $\Delta$ value to
small mass objects. One  clearly sees that for  $\Delta = 2$, negative
values of  the   density contrast  are selected,   showing  that shear
effects can actually lead to the collapse of under-dense regions.  For
$\Delta=0.5$,  the  selection  functions become steeper  and  converge
towards   an  Heavyside  function.   For  the   Zel'dovich  model, the
selection   function in the  limit  $\Delta \rightarrow  0$ leads to a
Heavyside step function at $\delta=3$, illustrating that this model is
only  a  first order approximation. For  the  AA model,  the selection
function converges  in  the limit $\Delta  \rightarrow  0$ towards the
Heavyside function  at   $\delta=\delta_c$.   The derivative   of  the
selection function $\partial S/\partial \Delta$ peaks at a value lower
than  $\delta_c$,  and this  peak   value tends towards  $\delta_c$ as
$\Delta  \rightarrow 0$.   Therefore, in the   AA approximation, where
objects are defined by the collapse of their first principal axis, the
mean initial   density contrast is  lower than  the standard spherical
threshold $\delta_c$.

We deduce now the universal mass function

\begin{equation} 
\label{eq20}
\Phi (\Delta)=-\frac{1}{F_0}\frac{1}{\sqrt{2\pi}}
\int _{-\infty} ^{+\infty}
e^{-\nu^{2}/2} \frac {\partial S}{\partial \Delta} d\nu
\end{equation}

\begin{figure}[httb]
\begin{picture}(9,9)
\put(0,0){
\psfig{file=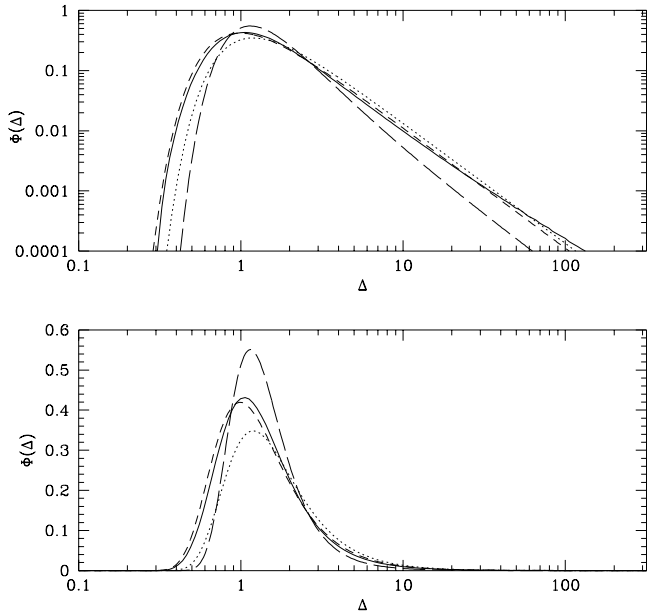,height=9.cm,width=9.cm}}
\end{picture}
\caption{
  Universal multiplicity functions  on linear and logarithmic  scales. 
  The solid  line corresponds to our dynamical  model  (Audit \& Alimi
  1996), the  dashed line  to the PS   mass function with  $\delta_c =
  1.4$, the dotted line to the standard  PS mass function ($\delta_c =
  1.686$) and the long--dashed line to the Zel'dovich approximation.}
\label{fmuniv}
\end{figure}

For the spherical dynamics $F_0=0.5$ since only 50\% of the regions of
space collapse  (those with a  positive   density contrast).  For  the
Zel'dovich and AA approximations, $F_0\simeq 0.92$ as all regions with
a  negative $\lambda_3$  collapse.  In  figure  (\ref{fmuniv}) we plot
$\Phi(\Delta)$ for the three different dynamics.  The bottom figure is
shown with a linear scale to allow an easy  examination of the maximum
amplitude of the mass functions.  The top one,  shown on a logarithmic
scale, is better adapted  to observing the  small mass (high $\Delta$)
slope and the high mass (small  $\Delta$) cut--off.  The mass function
computed with our dynamical model (full line) is very well fitted by a
standard PS  mass   function   with  a  critical density     parameter
$\delta_c=1.4$ (short--dashed line).  This  is easily explained by the
behavior   of  the  selection   function  previously  discussed, whose
effective threshold is lower  than the spherical threshold $\delta_c$. 
On the contrary, for the  Zel'dovich approximation, the resulting mass
function cannot be approximated by a PS function with another value of
$\delta_c$.  The high--mass cut--off is however located at lower mass,
as can be expected  from the selection  function which chooses regions
with an effective threshold higher than the standard value $\delta_c$.
All  mass functions  have  an exponential cut-off   at high mass and a
power law behavior at small mass. However,  this doesn't mean that the
resulting mass function can  always be fitted by  the PS formula, with
$\delta_c$ as fit parameter.

Moreover, we see  that the standard ($\delta_c  \simeq 1.686$) PS mass
function   (dotted line) and our  mass  function have roughly the same
slope  for small  mass  objects, although  we  predict more high  mass
objects than the standard PS.   For example, if  the power spectrum is
normalised with the condition  $\Delta(R=8 Mpc.h^{-1})=1$ and assuming
assume a critical Univers with $h=0.5$, we find that our mass function
contains $30\%$  more structures  of mass $\approx  10^{15}M_{\odot}$,
which is the typical mass of galaxy clusters.

%%%%%%%%%%%%%%%%%%%%%%%%%%%%%%%%%%
\section{Beyond Lagrangian Models}
%%%%%%%%%%%%%%%%%%%%%%%%%%%%%%%%%%

In  the  previous section we have  computed  the mass  function in the
framework  of the PS  formalism  using different Lagrangian  dynamical
models  to determine the  collapse epoch  of a  fluid element. In  all
cases, the collapse was defined as the epoch when the density contrast
becomes infinite.  It  is then impossible  to follow,  with Lagrangian
dynamics,   the evolution  of  a   fluid   element beyond this   first
singularity.  Moreover, if    one considers that the  fluid   elements
describe extended regions of space, this first singularity corresponds
to a  stage where  the  region has totally  collapsed along   a single
direction  and    reached  a  sheet--like  geometry.    The  resulting
``pancakes'' may, however, not be  the dense virialized halos that one
would expect  as  well defined objects in  the  density field.   It is
therefore  interesting to investigate what  happens when one waits for
the collapse of the fluid element along its second or third axis which
corresponds  to filamentary or   quasi-spherical objects respectively. 
The mass function deduced from the collapse  of the third axis will be
presented in greater detail, as it is, in our opinion, best suited for
dense virialized objects in the density field.

In section 2.3   and figure  (\ref{invtc}),  we  have seen that    the
dispersion  of the  collapse epoch  of  the first    axis, in the   AA
approximation, due to the shear on the second axis was of the order of
20\%.  This weak   dependence  suggests that,   as in the   Zel'dovich
approximation, the different directions   of the fluid element can  be
considered as collapsing independently.   This approximation allows to
artificially go beyond the first  singularity and to propose an ansatz
for the collapse epoch of the second and third axis in order.

We require that the collapse epoch of a fluid element along any of its
principal axis satisfies the following properties:

\begin{itemize}
\item  The collapse  epoch  along a given axis   is a function  of the
  corresponding initial shear  eigenvalue and  of the initial  density
  contrast.
\item  The scaling  properties   (eq.  [\ref{scalinglaw}])  should  be
  satisfied.
\item The spherical collapse epoch  should be recovered if the initial
  eigenvalue of the shear along this direction is zero.
\item The collapse epoch should be a monotonically increasing function
  of the shear.  A  positive (resp.  negative) shear eigenvalue should
  slow   down (resp.  accelerate)   the   collapse relative  to  the
  corresponding spherical model.
\end{itemize}

\noindent
These properties mean that the following formulae hold 

\begin{equation}
a_c(\delta,\sigma_i)=\frac{1}{\delta}a_c(1,\sigma_i/\delta)
                    =\frac{1}{\delta}\tilde{a}_c(\sigma_i/\delta) 
\end{equation} 

\noindent 
where   $\tilde{a}_c$   is a    monotonous increasing   function  and
$\tilde{a}_c(0)=\delta_c$.   Since $\tilde{a}_c$ is monotonous, it is
also bijective and  has an inverse,  $\tilde{\sigma}_c $,  which gives
the shear corresponding to  a given collapse  epoch in the $\delta=+1$
plane (note that there are, as in the  previous section, two functions
$\tilde{a}_c$,   one for the positive and  one   for the negative   density
contrasts).

%%%%%%%%%%%%%%%%%%%%%%%%%%%%%%%%%%%%%%%%%%%%%%%%%%%%%%%%%
\subsection{Mass Functions of Sheets, Filaments or Knots}
%%%%%%%%%%%%%%%%%%%%%%%%%%%%%%%%%%%%%%%%%%%%%%%%%%%%%%%%%

In order to push our investigations further we now propose a simple
ansatz for the collapse time along each axis.  We do not have an
underlying dynamical model to justify for the following approximation of
the collapse epoch of the different axes. These models are  a
phenomenological approach to studying qualitatively the influence of
considering the collapse along a given axis.

For $\delta > 0$ and $\sigma_i < 0$ as well as for $\delta < 0$ we
approximate by linear functions the results given in the previous
section

\begin{eqnarray}
\label{tc1}
1 + z_c &=& \delta/\delta_c-\alpha \sigma_{i} 
\qquad \mbox{ if } \delta > 0 \mbox{ and } \sigma_i <0\\
\label{tc2}
1+z_c&=&\alpha(\delta/3-\sigma_{i})
\;\;\, \mbox{ if } \delta < 0 
\end{eqnarray}

\begin{figure}[httb]
\begin{picture}(9.,4.5)
\put(-0.5,0){
\psfig{file=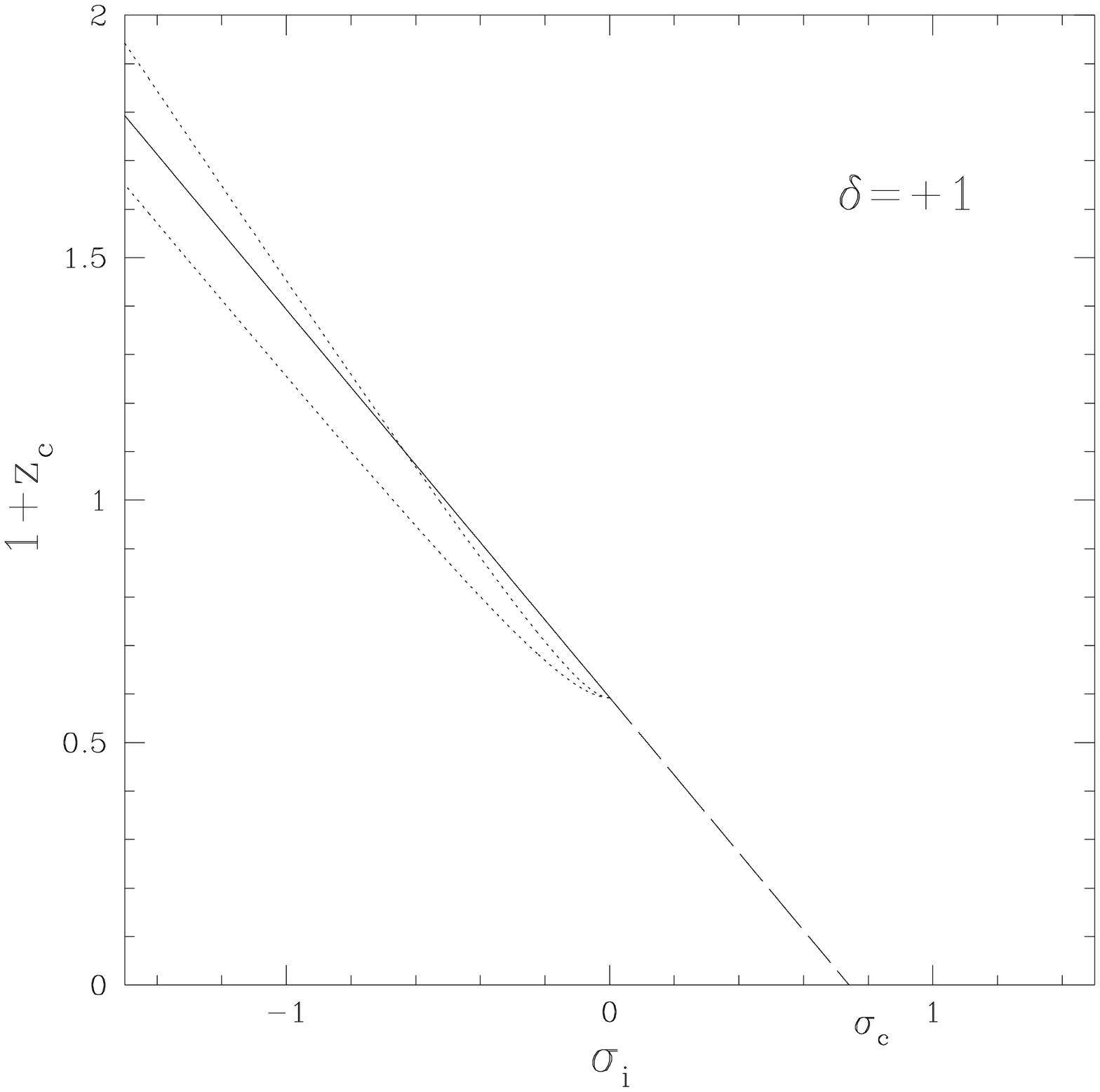,height=4.5cm,width=4.5cm}}
\put(4.,0){
\psfig{file=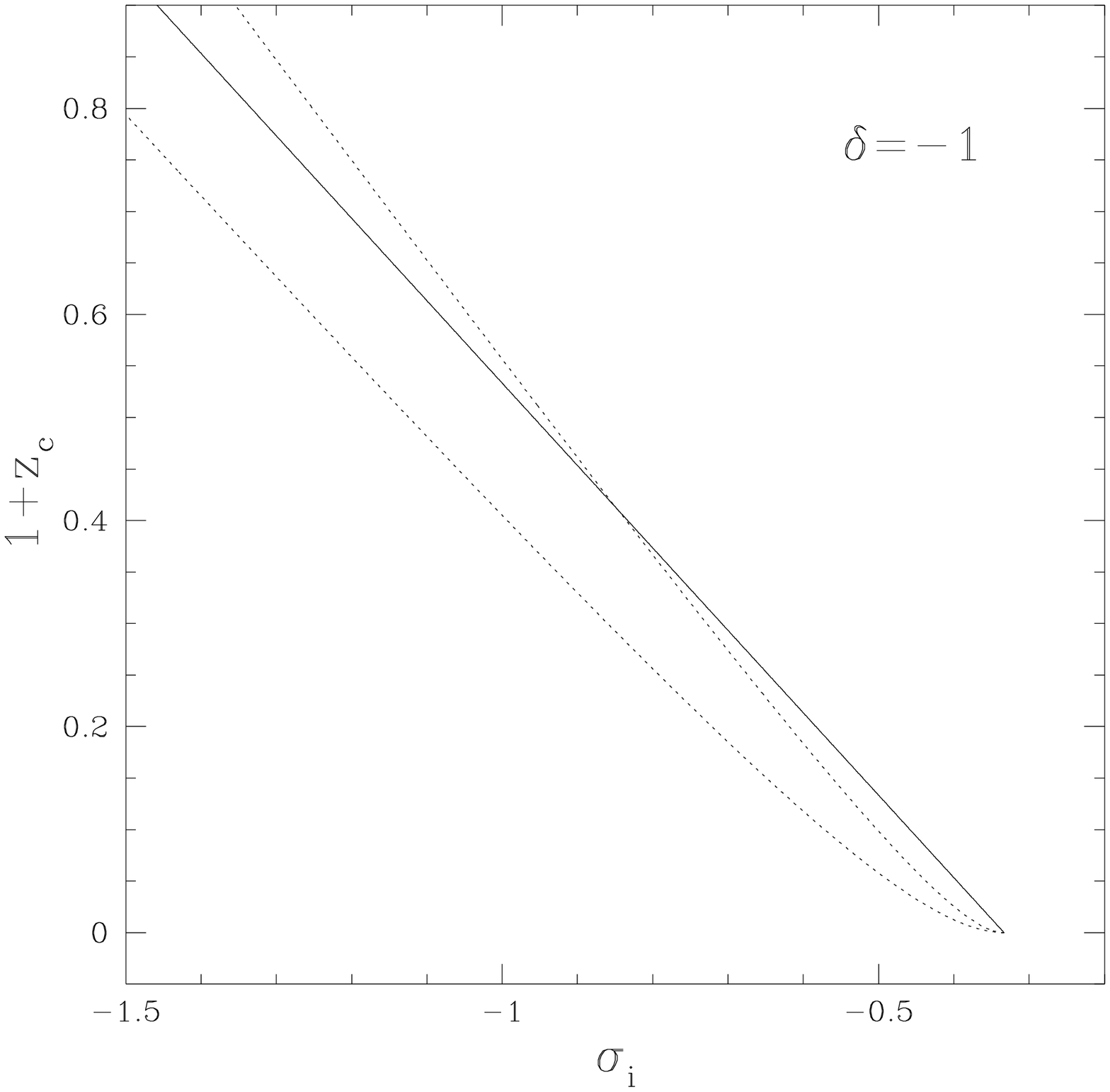,height=4.5cm,width=4.5cm}}
\end{picture}
\caption{The collapse redshift as a function of the eigenvalue of the 
  shear.  The ansatz defined by  Eqs.  (\ref{tc1}) and (\ref{tc2}) are
  shown by the full-line for the  part deduced from  the linear fit of
  the  results of  previous  section.  The dashed-line  represents the
  extrapolation  in the $\delta>0,  \sigma_i>0$  region  ($\alpha=0.8,
  \epsilon=1$) (Eq.  \ref{tc3}).  As a  reminder,  the results of  the
  previous section are shown as a dotted-line.}
\label{ans}
\end{figure}

\noindent 
With this model  we recover, for  $\delta  > 0$, the spherical  collapse
epoch for $\sigma_i=0$ and for $\delta < 0$.  The collapse time becomes
infinite for  $\sigma_i=\delta/3$.    The  slope $\alpha$  is   a free
parameter, it  has  to satisfy $0.8  \le   \alpha \le 1$   in order to
recover approximately the right collapse time for the first axis
(fig. \ref{ans}).

To compute the mass function  for any axis, we also  need to know  the
collapse epoch for $\delta > 0$  and $\sigma_i > 0$. Unfortunately, in
this case we do not have any dynamical model to determine the function
$\tilde{a}_c$.  However, to study the qualitative behavior of the mass
function we choose a simple ansatz for the collapse time given by

\begin{equation} 
1+z_c = \frac{\delta}{\delta_c}
(1-\frac{\sigma_i}{\sigma_c\delta})^{\epsilon}
\label{tc3}
\end{equation} 

\noindent
The spherical collapse  time is once again  found for vanishing  shear
The  shear  totally prevents the fluid   element from  collapsing when
$\sigma_i \ge  \sigma_c   \delta$.   Note that  for   large  values of
$\sigma_c$ we  find the spherical  model  everywhere.  $\epsilon$ is a
positive, free parameter which   determines the  global shape  of  the
function $\tilde{a}_c$.  This  parameterization  is somewhat arbitrary
but  satisfies the general   requirements   presented above.  We   now
compute the selection function, its derivative  and the universal mass
function for objects which collapse  along the first, second or  third
axis.

\subsubsection{First axis collapse}

\begin{figure}[httb]
\psfig{file=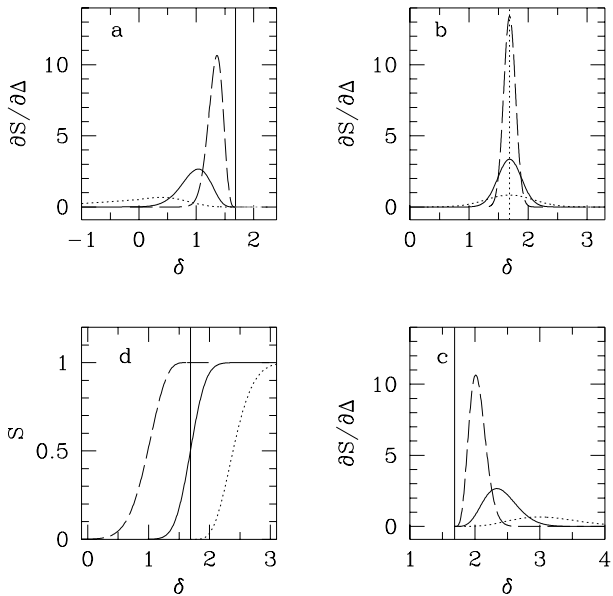,height=9cm}
\caption{Derivative of the selection function for $\Delta=0.5$, 
  $1$ and $2$ (long-dashed, solid  and dotted lines respectively)  and
  for first,  second  and  third axis  collapse   (fig.  a,   b and  c
  respectively).  Figure d shows the selection function for $\Delta=1$
  for the  collapse of the first, second  and third axes (long-dashed,
  solid and  dotted lines respectively).  In all  figures the vertical
  solid line indicates the position $\delta=\delta_c$.}
\label{sel3axis}
\end{figure}

This  case has been  studied in detail  in  the previous sections.  We
want  to show  here  that the results are    not strongly affected  in
neglecting the influence  of  $\sigma_2$.  Since $\sigma_3$ is  always
negative we use formulae (\ref{tc1})  and (\ref{tc2}) for the collapse
epoch.  We  have plotted   in figure (\ref{sel3axis})    the selection
function  and  its derivative   for the  case  $\alpha=0.8$   and  for
different values of $\Delta$.  The  difference with the results of the
previous sections is  always less than a few  percent.  As was already
mentioned before, the selection function is equal to unity for $\delta
> \delta_c$   and  continuously  decreases  towards  zero when $\delta
\rightarrow -\infty$.  Its  derivative sharpens  and  peaks at a value
which goes  towards $\delta_c$  (while remaining smaller)  as $\Delta$
decreases.    The resulting   mass   function  is plotted  on   figure
(\ref{fmuni3axis})  and  always differs by less  than  10\% to the one
obtained in the previous section.

\subsubsection{Second Axis Collapse}

The  collapse epoch of  the second axis is  a function of $\delta$ and
$\sigma_2$.  In the last  section, $\sigma_3$ varies between $-\infty$
and $0$.  Now, $\sigma_2$ belongs  to the domain $]-\infty,+\infty[$.  
Therefore  we use for   the collapse epoch, the formulae  (\ref{tc1}),
(\ref{tc2})  and    (\ref{tc3}).     We  have     plotted   in  figure
(\ref{sel3axis}) the selection   function and its derivative  for  the
case  $\alpha=0.8$, $\epsilon=1$  and $\sigma_c=1/\alpha\delta_c$  (we
have just extrapolated formula (\ref{tc1})  for the domain $\sigma_2 >
0$).   In this case the selection  function increases from zero to one
when $\delta$  goes from  $-\infty$   to  $+\infty$ and  is  equal  to
one-half  for $\delta=\delta_c$  (all  the points with  $\mu_2 \le  0$
(50\%) which collapse  faster than the  spherical model are selected). 
As before, a  fraction of the  fluid elements with $\delta < \delta_c$
are selected, but contrary  to what happened  in the previous  case, a
fraction of the fluid elements with  $\delta > \delta_c$ are unable to
collapse  because of shear effects.  The   derivative of the selection
function  peaks  at around $\delta_c$  (exactly at  $\delta_c$ for our
model) and  has a  width which tends   towards zero in the  limits  of
vanishing $\Delta$.

The  resulting mass function is  plotted in figure (\ref{fmuni3axis}). 
It is now quite  close to the  standard PS mass function,  which means
that the  number of objects with  $\Delta \lesssim 1$ has considerably
decreased compared to the first axis  collapse.  This fact is not very
surprising because  in this case, the  shear can either  slow down the
collapse for $\sigma_2 > 0$ or accelerate it for $\sigma_2 < 0$. These
two effects     roughly   cancel each other,     and  one   obtains  a
quasi--spherical model.

\subsubsection{Third Axis Collapse}

\begin{figure}[httb]
\psfig{file=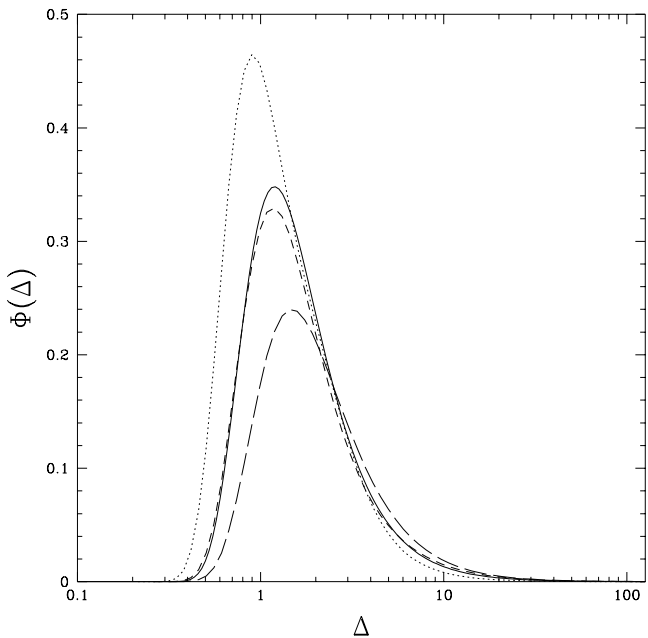,height=9cm}
\caption{Mass function obtained for  first, second and third axes
  collapse (dotted, dashed and  long-dashed lines). The solid line  is
  the classical ($\delta_c=1.686$) Press and Schechter mass function.}
\label{fmuni3axis}
\end{figure}

We  now present in  greater detail   the  mass function obtained  for
objects   that collapse  along their  third  axis.   In this  case the
collapse  epoch  is  a  function  of  $\delta$ and  $\sigma_1  \in [0,
+\infty[$.  It   is possible   to derive analytically    the selection
function, its  derivative  and the  corresponding  mass function for a
given $\tilde{a}_c$.   After  some algebra,   we obtain 

\begin{equation} 
S_{\tilde{a}_c}(\nu\Delta,\Delta)= 
\int_{0}^{\nu\tilde{\sigma}_c(\nu\Delta)} P_{\mu_1}d\mu_1
\end{equation} 

\noindent
Taking the derivative of this equation with respect to
$\Delta$ yields

\begin{equation}
\label{eq26} 
\frac{\partial S_{\tilde{a}_c}}{\partial \Delta}(\nu\Delta,\Delta)= 
P_{\mu_1}
\left(\nu\tilde{\sigma}_c(\nu\Delta)\right)
\nu^2
\dot{\tilde{\sigma}}_c(\nu\Delta) 
\end{equation} 

\noindent 
where $ \dot{\tilde{\sigma}}  _c $  symbolizes the derivative  of the
function ${\tilde{\sigma}} _c$.     Finally, using the    variable $x$
defined   by   $x=\tilde{\sigma}_c(\nu\Delta)$,   the universal   mass
function can be expressed as
 
\begin{equation} 
\!\!\!\!\!\!\!\!
\Phi(\Delta) = \frac{F_0^{-1}}{\sqrt{2\pi}} \int_{0}^{\sigma_c} 
\!\!\!\! e^{-\frac{{\tilde{a}_c}^2(x)}{ 2\Delta^2}}
P_{\mu_1} \left( \frac{x \tilde{a}_c(x)}{\Delta} \right)
\frac{{\tilde{a}_c}^2(x)}{\Delta^3}dx
\label{fm3axis}
\end{equation} 

\noindent 
where  $\sigma_c \in  ]0,+\infty]$ is   the  value  of  $x$  at  which
$\tilde{a}_c$ becomes infinite.  A similar  calculation can be carried
out   for all axes.   The  above formula  is  a   general relation
between   the mass  function   and the  collapse   time defined by the
underlying dynamical model.

For the third axis, we assume that the  collapse epoch is now given by
equation (\ref{tc3}) and we use the same  parameters as for the second
axis.  The selection function (Fig.   [\ref{sel3axis}]) tends to  zero
as $\delta  \rightarrow \delta_c$ and to  unity as $\delta \rightarrow
+\infty$. This reflects the fact that a fraction of the fluid elements
with $\delta \ge  \delta_c$  are  unable  to  collapse as   the  shear
inhibits the collapse.

The derivative of  the selection function behaves symmetrically  about
the line $\delta=\delta_c$ compared to  the one obtained for the first
axis. Its derivative sharpens and peaks at a value which tends towards
$\delta_c$  (while remaining larger)   as $\Delta$ decreases.  In  the
class  of models  with  $\epsilon = 1$,  it  is possible to derive  an
analytical expression for these qualitative  features.  We compute the
mean  (which  is  close to  the peak  value)  and the  variance (which
corresponds to the  width)  of the   function $\partial S   / \partial
\Delta$. We obtain for the mean

\begin{equation}
\bar \delta = \delta_c
+\frac{3}{5}\sqrt{\frac{5}{2\pi}}\frac{\Delta}{\sigma_c}
\simeq \delta_c + 0.54 \frac{\Delta}{\sigma_c}
\end{equation}

\noindent
and for the variance

\begin{equation}
\Sigma = \frac{1}{3}\sqrt{\frac{29\pi-81}{10\pi}}\frac{\Delta}{\sigma_c}
\simeq 0.19 \frac{\Delta}{\sigma_c}
\end{equation} 

\noindent
In these last two equations the width and the peak position are linear
functions of $\Delta$. Note also  that as  $\Delta \rightarrow 0$,  we
get $\bar   \delta =\delta_c$ and $\Sigma  =  0$, as  in  the spherical
model.

The  mass  function can also  be  derived  analytically. Using formula
(\ref{fm3axis}), we obtain

\begin{equation}
\Phi(\Delta) = \frac{F_0^{-1}}{\sqrt{2\pi}}\frac{\delta_c}{\Delta^2}
 \int_{0}^{+\infty} \!\!\!\! e^{-\frac{1}{2}\left(
 \frac{\delta_c}{\Delta}+\frac{\mu}{\sigma_c}\right)^2}
  P_{\mu_1}(\mu)d\mu
\end{equation}

The   mass  function  now  has two   ``dynamical  parameters'', namely
$\delta_c$  and $\sigma_c$.  The  first one describes the influence of
the initial density contrast on the  collapse epoch and the second one
gives the degree  of inhibition of the  collapse by the shear and  the
tidal field.  The  spherical mass function  is recovered in the limits
$\sigma_c \rightarrow +\infty$ and for small values of $\sigma_c$ most
of the fluid elements are unable to collapse because of shear effects.
This mass function for $\sigma_c=0.74$ and $\delta_c=1.686$ is plotted
in  figure (\ref{fmuni3axis}) as a long--dashed  line.   The number of
objects with $\Delta \lesssim 1$ has  again greatly decreased compared
to the second  axis  collapse.  As could  be  expected, since now  the
shear always  slows down the  collapse, the mass  function is now much
below the standard PS prediction.

In the next section  we investigate the  generic behavior of  the mass
function   obtained in   this  section (Eq.   [\ref{fm3axis}]) in  the
low--mass regime.

%%%%%%%%%%%%%%%%%%%%%%%%%%%%%%%%%%%%%%%%%%%%%%%%%%%%
\subsection{Low Mass Behavior of the Mass Functions}
%%%%%%%%%%%%%%%%%%%%%%%%%%%%%%%%%%%%%%%%%%%%%%%%%%%%

In the previous section, we postulated a specific form for the
collapse epoch of a given fluid element, as a function of $\delta$ and
the shear along the third axis.  In this section, we study in a more
general way the influence of the shear on the formation of low mass
objects.  We consider that the function $\tilde{a}_c$ has the
following asymptotic behavior when $x \rightarrow \sigma_c$

\begin{equation}
\label{eq31} 
\tilde{a}_c @>>{\sigma_c}>\left(\sigma_c - x\right)^{-\epsilon}
\end{equation} 

\noindent
where $\epsilon > 0$. This parameterization is very general, and makes
no assumptions on the general  shape of the function $\tilde{a}_c$. We
calculate analytically the low--mass  (high--$\Delta$) behavior of the
mass function as  a function of $\epsilon$.   After some algebra (cf.  
Appendix B), we obtain two  characteristic regimes, valid in the limit
$\Delta \gg 1$

\begin{alignat}{2}
\label{li1}
\Phi & \propto  \Delta^{-\left(\frac{\epsilon+1}{\epsilon}\right)}& 
\quad & \mbox{ if } \epsilon \ge \frac{1}{6}\\
\label{li2}
\Phi & \propto  \Delta^{-7} & \quad
& \mbox{ if } \epsilon \le \frac{1}{6} 
\end{alignat} 

\noindent
Note that  for the case  $\epsilon=1$, the  low--mass  behavior is the
same  as  for the standard  Press   \& Schechter  mass function. The
domain  $\epsilon >   1$  corresponds to   shallower slopes  than  the
standard   PS mass  function.   However, the  Zel'dovich  approximation
corresponds  to $\epsilon=1$.  Therefore,  if one considers  that this
first   order   approximation overestimates   the  collapse time, this
imposes that

\begin{equation} 
0 < \epsilon < 1
\end{equation} 

\begin{figure}[httb]
\psfig{file=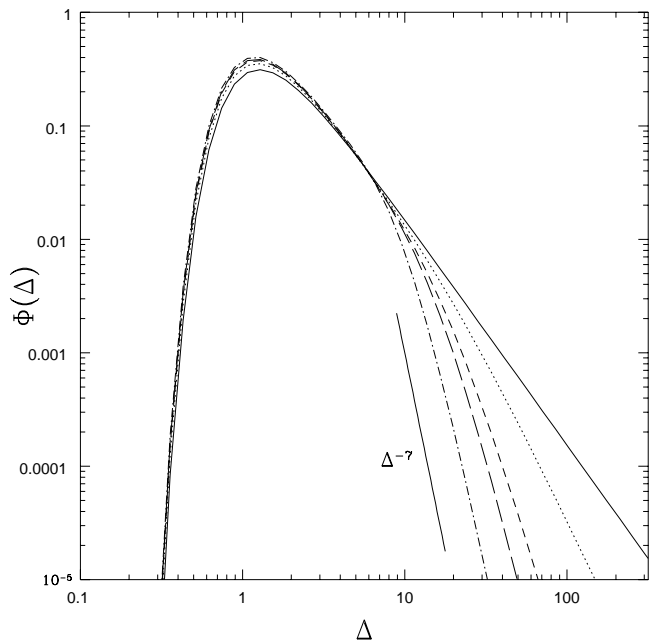,height=9cm}
\caption{Mass function obtained for the third axis collapse and for
  $\epsilon$=1, 1/2,  1/4,  1/6  and  1/100  (solid,  dotted,  dashed,
  long-dashed and dotted-dashed lines respectively). Note that for
  $\epsilon \le 1/6$ the slope at low mass remains equal to $-7$.} 
\label{lowmass}
\end{figure}

We plot in figure (\ref{lowmass}) the mass functions we obtained using
the collapse epoch from formula (\ref{tc3}) for different values of
$\epsilon \le 1$.  We set the parameter $\sigma_c=3$ for each curve in
figure (\ref{lowmass}).  This parameter determines the point where the
change of   slope  occurs in  the  mass  function (in  our  case, this
corresponds to $\Delta \simeq 5$).

This result is extremely important as it  shows that shear effects can
easily modify the low--mass behavior of the mass function.  The regime
$\epsilon \le 1/6$ is of primary importance, because it is independent
of  the   specific form  of   the  function  $\tilde{a}_c$, which   is
unfortunately an unknown quantity. In this case, for a power law power
spectrum,  one  obtains  in the  low--mass  end of  the  mass function
$\Phi(M) \propto   M^{n+1}$ where  $n$  is  the   index  of  the power
spectrum.  Assuming a constant mass--to--light  ratio, this leads to a
faint--end slope of  the luminosity function  $\alpha=n+1$.  Note that
the corresponding power  law for the  standard PS mass function  would
have been $\alpha = -2+(n+3)/6$.   The observed luminosity function of
field  galaxies exhibits this  power  law behavior at  low--luminosity
with  $\alpha \simeq -1$ (Loveday et   al.  1994; Efstathiou, Ellis \&
Peterson 1988). It has been previously argued (Blanchard et al.  1992)
that for a CDM power  spectrum with $n \simeq  -2$ at these scales, it
is impossible to  reconcile the PS  formalism  ($\alpha \simeq -1.83$)
with  the observed luminosity  function without  imposing some ad--hoc
procedures, such as  a  variable mass--to--light ratio, or  some other
biasing mechanism due to the dissipative physics of  the baryonic gas. 
Here, we recover  $\alpha \simeq -1$ with a  power spectrum $n  \simeq
-2$, using  only  gravitational  processes,  and  more   precisely  by
considering  the effect  of   the shear.   We   claim here that  the a
CDM--like power spectrum is able to reproduce the low--mass end of the
luminosity function, when the  shear and  tidal effects are  correctly
taken into account.  Note that  the PS formalism  is better suited for
field galaxies  than for galaxies  in clusters where tidal effects are
much stronger.  With this formalism,  the  clusters are considered  as
single objects whose internal structure can not be described.

%%%%%%%%%%%%%%%%%%%%
\section{Conclusion}
%%%%%%%%%%%%%%%%%%%%

In this paper, we extend the PS formalism  to more complex, and perhaps
more realistic dynamical models than the standard spherical model.  We
study the  influence  of the  shear  and tidal  effects  on  the  mass
function of cosmic structures.

The Lagrangian evolution of a given halo from the initial volume up to
the final dense  and virialized   object,  can be divided into   three
stages: the collapse of  the first, second  and third axes. Each stage
corresponds  to  a certain geometry  of  the halo formed (sheet--like,
filamentary and  quasi--spherical)  and  also to a   certain  collapse
epoch. In the Lagrangian formalism, this collapse epoch depends purely
on the initial shear and density contrast.  The mass function of these
halos is then obtained from the collapse epoch and from the statistics
of the initial parameters.

We investigate in this  paper the influence of  the initial shear  on
the dynamics of the forming halos. In the case of sheet--like objects,
the Lagrangian theory  of gravitational dynamics  allows us to compute
the  collapse epoch  of  the first  principal  axis  as a  function of
$\delta$ and $\sigma_3$  and $\sigma_2$, the initial density  contrast
and  two  eigenvalues of the  initial  shear tensor.  We use  for that
purpose  a new approximation  presented by Audit \&  Alimi (1996).  In
this case we obtain a mass function which is well  fitted by a PS mass
function,  with $\delta_c=1.4$. This result  is  explained by the fact
that shear and tidal effects always accelerate the collapse of a fluid
element  along   its first principal  axes. We   therefore obtain more
high--mass objects than the standard PS mass function (see also Monaco
1995).

We then investigate  the mass function  of objects that collapse along
their second and  third    principal axis.  We considere    simple and
general models since  Lagrangian  theory  is  unable to describe   the
evolution of a fluid element beyond the collapse of the first axis. We
find that the  resulting mass function in the  case of the second axis
collapse is very  similar to the standard  PS mass function ($\delta_c
\simeq 1.686$).

Finally, we derive analytically a new formula for the mass function of
objects resulting from the collapse of their third axis.  This depends
on two parameters,  namely a density threshold  $\delta_c$ and a shear
threshold $\sigma_c$.  We recover the  standard PS formula in the case
$\sigma_c \rightarrow +\infty$.  For  values of  $\sigma_c$ comparable
to unity, the peak of the multiplicity function is lower than the peak
obtained by the standard PS approach.  This  could explain why such an
effect has  been detected in numerical simulations  (Efstathiou et al. 
1988).   We also derive analytically  the low--mass regime of the mass
function computed  in  a wide class  of  models.  We  show that  it is
possible  to  obtain a low--mass slope   very  close to the faint--end
slope of the luminosity function of field galaxies.

In all the dynamical models we use, the  resulting mass functions show
common  features  such  as a  high--mass exponential    cut--off and a
low--mass power  law.  The  exact  shape of the  multiplicity function
around  the peak   is,  however, directly  related  to the  underlying
dynamical model. This has to be tested in numerical simulations, where
the mass  function should correspond  to a dynamical model that drives
the collapse of every halo (at least in a statistical sense).  We
postpone this numerical work to a companion paper.

\appendix

\section{Collapse Epoch using our Dynamical Model}
 
In this appendix we  present the collapse   time obtained from  the AA
dynamics.  The following formulae were obtained  by fitting a thousand
points for $\delta  >  0$ and $\delta  <0$.  The fits are  accurate to
about 1\%.

For $\delta > 0$, we obtain

$$
{\cal T}^{+} = 1 + z_{c} =
\frac{x^{3}-1}{7}f^{+}  +  \frac{8-x^3}{7}f^{-}
$$

\noindent
where $x \in [1,2]$ is defined by

$$
x = \left( 1 + \frac{1}{3} \left(1+2\frac{\sigma_2}{\sigma_3}\right)^2
    \right)^{1/2}
$$

\noindent
The functions $f^+$ and  $f^-$ depend only on $\sigma_3/\delta$,  and
are  proportional to   $\delta$,  according to the  scaling  property
derived in section 2.

$$
f^+ = \delta \left( \frac{1}{\delta_c} - \frac{\sigma_3}{\delta} - 0.2
    (1 - g^+) \right)
$$
$$
g^+ = \left( 1 - 10\sigma_3 / \delta \right)^{-1/2}
$$
$$
f^- = \delta \left( \frac{1}{\delta_c} - 0.81\frac{\sigma_3}{\delta} - 0.18
    (1 - g^-) \right)
$$
$$
g^- = \left( 1 - 4.5\sigma_3/ \delta \right)^{-1}
$$

\noindent
Note that the spherical model is recovered for vanishing shear. The
Zel'dovich solution, which is exact when $\sigma_{3}=-2\delta/3$ and
$\sigma_{2}=-\sigma_{3}/2$, is also recovered.

For $\delta < 0$, we have

$$
{\cal T}^- =  1 + z_c  = 
\frac{x^{3a}-1}{8^a - 1}f^+  +  \frac{8-x^{3a}}{8^a - 1}f^-
$$

\noindent
where $x \in [1,2]$ is the same as for $\delta > 0$ and $a$ is given by

$$
a = \left(1+ \frac{0.5}{3\sigma_3/ \delta - 1} \right) ^{1/2}
$$
\noindent
The functions $f^+$ and  $f^-$ are now given by
$$
f^+ = - \delta \left( \left(\frac{\sigma_3}{\delta}- \frac{1}{3}\right)
 - 0.2 (1 - g^+) \right)
$$
$$
g^+ = \left( 1 + 15\left(\sigma_3 / \delta - 1/3 \right)
 \right)^{-1/3}
$$
$$
f^- = - \delta \left( 0.81\left(\frac{\sigma_3}{\delta}-\frac{1}{3} \right) 
- 0.18 (1 - g^-) \right)
$$
$$
g^- = \left( 1 + 4.5\left(\sigma_3/ \delta - 1/3 \right)
 \right)^{-1}
$$

\noindent
When the shear is  much larger than  the density contrast  both ${\cal
  T}^+$ and ${\cal  T}^-$ converge  toward the same   value as 
expected.

\section{Computation of the low-mass behavior of the mass function 
  for the third axis collapse}

We  present in this appendix  the derivation of  the low-mass behavior
presented   in  section   (3.2).   From   equations (\ref{eq20})   and
(\ref{eq26}) the universal mass function can be written as

\begin{equation} 
\Phi(\Delta) \propto \int_{\delta_c/\Delta}^{+\infty} e^{-\nu^{2}/2} 
              P_{\mu_{1}}(\nu \tilde{\sigma_c}(\nu\Delta))\nu^{2} 
              \dot{\tilde{\sigma_c}}(\nu\Delta) d\nu
\end{equation} 
 
We want to  study the behavior of the  previous integral for  the case
where $\Delta \rightarrow \infty$.  The collapse time has the behavior
given by equation (\ref{eq31}) which allows to calculate the following
limits

\begin{equation}  
\label{l1}
\tilde{\sigma}_c(x) @>>{\infty}> \sigma_c  \: \mbox{ and } \:
\dot{\tilde{\sigma}}_c(x) @>>{\infty}> 
\frac{1}{\epsilon}x^{-(\epsilon+1)/\epsilon}
\end{equation} 

We then divide into two the integral giving the mass function 
\begin{eqnarray}
& \cal{I}_1=\int_{1}^{+\infty} e^{-\nu^{2}/2} 
              P_{\mu_{1}}(\nu \tilde{\sigma_c}(\nu\Delta))\nu^{2} 
              \dot{\tilde{\sigma_c}}(\nu\Delta) d\nu \\
& \cal{I}_2=\int_{\delta_c/\Delta}^{1} e^{-\nu^{2}/2} 
              P_{\mu_{1}}(\nu \tilde{\sigma_c}(\nu\Delta))\nu^{2} 
              \dot{\tilde{\sigma_c}}(\nu\Delta) d\nu 
\end{eqnarray} 
Using the limits given above, we have for large values of $\Delta$
\begin{equation} 
\cal{I}_1 \sim \int_{1}^{+\infty} e^{-\nu^{2}/2} 
              P_{\mu_{1}}(\nu\sigma_c)\nu^{2} 
              \frac{1}{\epsilon}(\nu\Delta)^{-(\epsilon+1)/\epsilon} d\nu 
\end{equation} 
Therefore, the asymptotic behavior of $\cal{I}_1$ when  $\Delta \rightarrow
\infty$ is 
\begin{equation} 
\label{c1}
\cal{I}_1 \sim \Delta^{-\frac{\epsilon+1}{\epsilon}}
\end{equation} 
Changing back  from $\nu$ to  $\delta$ and  using  the limits given in
equations (\ref{l1}), $\cal{I}_2$ can be written as
\begin{eqnarray} 
\cal{I}_2& = &\int_{\delta_c}^{\Delta} e^{-\delta^{2}/2\Delta^{2}} 
              P_{\mu_{1}}(\frac{\delta}{\Delta}
              \tilde{\sigma_c}(\delta))(\frac{\delta}{\Delta})^{2} 
              \dot{\tilde{\sigma_c}}(\delta) \frac{d\delta}{\Delta}\\
& = &\int_{\delta_c}^{\Delta} \left(\sum_{n=2}^{\infty} f_{n}\right)d\delta
\end{eqnarray}
where $f_{n}=\alpha_{n}\frac{\delta^{2n+1-1/\epsilon}}{\Delta^{2n+3}}$
and {\bf $\alpha_{n}$}   are numerical constants.   ($P_{\mu_{1}}$ and
the exponential have been developed in series).

If $\epsilon \geq 1/2(n+1)$, $\int f_{n} d\delta$ diverges and we have
\begin{equation}
\label{c2} 
\int_{\delta_c}^{\Delta} f_n d\delta \sim \Delta^{-\frac{\epsilon+1}{\epsilon}}
\end{equation} 

If $\epsilon < 1/2(n+1) \leq 1/6$, $\int f_{n} d\delta$ converges and 
therefore
\begin{equation}
\label{c3} 
\int_{\delta_c}^{\Delta} f_n d\delta \sim \Delta^{-2n+3} \leq \Delta^{-7}
\end{equation} 

Keeping only the leading terms from equations (\ref{c1}), (\ref{c2})
and  (\ref{c3}) gives the asymptotic  behavior  of $\Phi(\Delta)$ (Eqs. 
(\ref{li1}) and (\ref{li2})).


\begin{thebibliography}{}

\bibitem[Audit \& Alimi 1996]{AA96} Audit, A. \& Alimi, J-M.,  1996,
  A\&A, {\bf 315}, 11
  
\bibitem[Barbosa et al.  1996]{BBBO96} Barbosa D., Bartlett J.,
  Blanchard A.\& Oukbir J., 1996, A\&A, in Press
  
\bibitem[Bertschinger \& Hamilton 1994]{BH}Bertschinger, E. \&
  Hamilton, A.J.S., 1994 ApJ, {\bf 435}, 1

\bibitem[Blanchard et al.  1992]{BVGM92}Blanchard A., Valls-Gabaud  D. 
  \& Mamon G., 1992, A\&A, {\bf 264}, 365

\bibitem[Bond et al. 1991a]{B91a} Bond J.R., Cole S., Matarrese S.,
  Moscardini L., 1991,  MNRAS, {\bf 268}, 996

\bibitem[Bond et al. 1991b]{B91b} Bond J.R., Cole S., Efstathiou, G.,
  Kaiser, N., 1991, ApJ, {\bf 379}, 440

\bibitem[Calberg \& Couchman 1989]{CC89}Calberg, R.G. \& Couchman,
  H.M.P., 1989, ApJ, {\bf 340}, 47

\bibitem[Cole 1991]{C91} Cole S., 1991,  ApJ, {\bf 367}, 45

\bibitem[Cole \& Kaiser 1988]{CK88}Cole, S. \& Kaiser, N., 1988,
  MNRAS, {\bf 233}, 637

\bibitem[Doroshkevich 1970]{doro} Doroshkevich, A.G., 1970, Astrofizika
  {\bf 6}, 581 

\bibitem[Efstatiou et al. 1988]{EFWD88}Efstathiou, G.,
  Frenk, C.S., White, S.D.M. \& Davis, M., 1988, MNRAS, {\bf 235}, 715  

\bibitem[Efstatiou, Ellis \& Peterson 1988]{EEP88}Efstathiou, G.,
  Ellis, R.S. \& Peterson, B.A., 1988, MNRAS, {\bf 232}, 431

\bibitem[Eke et al. 1996]{Eke96} Eke, V.,
  Cole, S. \& Frenk, C., 1996, MNRAS, {\bf 283}, 263
  
\bibitem[Efstatiou \&  Rees 1988]{ER88}Efstathiou,  G., \& Rees, M.J.,
  1988, MNRAS, {\bf 230}, 5

\bibitem[Kofman \& Pogosyan 1995]{kof}  Kofman L. \& Pogosyan D.,
  1995, ApJ {\bf 442},30

\bibitem[Lacey \& Cole 1994]{LC94}Lacey, C. and Cole, S. 1994, MNRAS,
  {\bf 271}, 676

\bibitem[Lacey \& Cole 1993]{LC93}Lacey, C. and Cole, S. 1993, MNRAS,
  {\bf 262}, 627
  
\bibitem[Loveday  et   al.   1992]{L92}Loveday,   J., Peterson,  B.A.,
  Efstathiou, G. \&  Maddox, S.J., 1992, ApJ, {\bf 390}, 338

\bibitem[Monaco 1995]{M95}  Monaco, P., 1995,  ApJ, {\bf 447}, 23

\bibitem[Oukbir \& Blanchard 1992]{OB92} Oukbir, J. \& Blanchard, A.,
  1992, A\&A, {\bf 262}, 210

\bibitem[Peacock \& Heavens 1990]{PH90}Peacock J.A. \& Heavens A.F.,
  1990, MNRAS, {\bf 243}, 133
  
\bibitem[Peebles  1980]{P80}Peebles,  P.J.E.,  1980, ``The Large-Scale
  Structures of the Universe.'', (Princeton University Press)

\bibitem[Press \& Schechter 1974]{PS74} Press, W.H. \& Schechter, P., 1974,
  ApJ, {\bf 188}, 425

\bibitem[Van de Weygaert \& Babul 1994]{vdW94} Van de Weygaert, R.
  \& Babul, A. 1994, ApJ, {\bf 425}, L59

\bibitem[Viana \& Liddle 1996]{Via96} Viana, P.
  \& Liddle, R. 1996, MNRAS, {\bf 281}, 323

\bibitem[Zeldovich 1970]{Z70}Zel'dovich, Ya.B., 1970, A\&A.,
  {\bf 5}, 84

\bibitem[White \& Frenk 1991]{WK91}White, S.D.M. \& Frenk, C.S., 1991,
  ApJ, {\bf 379}, 25

\end{thebibliography}
\end{document}